\documentclass[dvips]{arxstspdf}
\usepackage{flushend}
\usepackage{stfloats}
\volume{26}
\issue{1}
\pubyear{2011}
\firstpage{10}
\lastpage{11}
\doi{10.1214/11-STS337C}
\referstodoi{10.1214/10-STS337}

\begin{document}
\begin{frontmatter}

\title{Bayesian Statistical Pragmatism}
\runtitle{Discussion}
\pdftitle{Bayesian Statistical Pragmatism. Discussion of Statistical Inference: The Big Picture by R. E. Kass}

\begin{aug}
\author{\fnms{Andrew} \snm{Gelman}\corref{}\ead[label=e1]{gelman@stat.columbia.edu}}

\runauthor{A. Gelman}

\affiliation{Columbia University}

\address{Andrew Gelman is Professor, Department of Statistics and Department of Political Science,
Columbia University, New York, New York 10027, USA \printead{e1}.}

\end{aug}

% ABSTRACT

% KEYWORDS

\end{frontmatter}

I agree with Rob Kass' point that we can and should make use of
statistical methods developed under different philosophies, and I am
happy to take the opportunity to elaborate on some of his arguments.

\section*{Foundations of Probability}

Kass describes probability theory
as anchored upon physical randomization (coin flips, die rolls and the
like) but being useful more generally as a mathematical model. I
completely agree but would also add another anchoring point:
\textit{calibration.} Calibration of probability assessments is an
objective, not subjective process, although some subjectivity (or
scientific judgment) is necessarily involved in the choice of events
used in the calibration. In that way, Bayesian probability calibration
is closely connected to frequentist probability statements, in that both
are conditional on ``reference sets'' of comparable events. We discuss
these issues further in Chapter 1 of \textit{Bayesian Data Analysis},
featuring examples from sports betting and record linkage.

\section*{Confidence Intervals and Hypothesis Tests}

I agree with Kass
that confidence and statistical significance are ``valuable inferential
tools.'' They are treated differently in classical and Bayesian
statistics, however. In the Neyman--Pearson theory of inference,
confidence and statistical significance are two sides of the same coin,
with a confidence interval being the set of parameter values not
rejected by a~significance test. Unfortunately, this approach falls
apart (or, at the very least, is extremely difficult) in problems with
high-dimensional parameter spaces that are characteristic of my own
applied work in social science and environmental health.

In a modern Bayesian approach, confidence intervals and hypothesis
testing are both important but are \textit{not} isomorphic; they
represent two different steps of inference. Confidence statements, or
posterior intervals, are summaries of inference about parameters
conditional on an assumed model. Hypothesis testing---or, more
generally, model checking---is the process of comparing observed data to
replications under the model if it were true. Statistically significance
in a hypothesis test corresponds to some aspect of the data which would
be unexpected under the model. For Bayesians, as for other statistical
researchers, both these steps of inferences are important: we want to
make use of the mathematics of probability to make conditionally valid
statements about unobserved quantities, and we also want to make use of
this same probability theory to reveal areas in which our models do not
fit the data.

\section*{Sampling}

Kass discusses the role of sampling as a model for
understanding statistical inference. But sampling is more than a
metaphor; it is crucial in many aspects of statistics. This is evident
in analysis of public opinion and health, where analyses rely on
random-sample national surveys, and in environmental statistics, where
continuous physical variables are studied using space-time samples. But
even in areas where sampling is less apparent, it can be important.
Consider medical experiments, where the object invariably is inference
for the general population, not mere\-ly for the patients in the study.
Similarly, the goal of Kass and his colleagues in their neuroscience
research is to learn about general aspects of human and animal brains,
not merely to study the particular creatures on which they have data.
Ultimately, \textit{sample} is just another word for \textit{subset},
and in both Bayesian and classical inference, appropriate generalization
from sample to population depends on a~model for the sampling or
selection process. I have no problem with Kass' use of sampling as a
framework for inference, and I think this will work even better if he
emphasizes the generalization from real samples to real
populations---not just mathematical constructs---that are central to so
much of our applied inferences.

\section*{Subjectivity and Belief}

The only two statements in Kass'
article that I clearly disagree with are the following two claims: ``the
only solid foundation for Bayesianism is subjective,'' and ``the most
fundamental belief of any scientist is that the theoretical and real
worlds are aligned.'' I will discuss the two statements in turn.

Claims of the subjectivity of Bayesian inference have been much debated,
and I am under no illusion that I can resolve them here. But I will
repeat my point made at the outset of this discussion that Bayesian
probability, like frequentist probability, is except in the simplest of
examples a model-based activity that is mathematically anchored by
physical randomization at one end and calibration to a~reference set at
the other. I will also repeat the familiar, but true,
argument\footnote{As a friend remarked to me in tenth-grade English
class, ``I don't know why they don't want us to use clich\'{e}s. These
sayings are clich\'{e}s because they're true!''} that most of the power
of a Bayesian inference typically comes from the likelihood, not the
prior, and a person who is really worried about subjective
model-building might profitably spend more effort thinking about
assumptions inherent in additive models, logistic regressions,
proportional hazards models, and the like. Even the Wilcoxon test is
based on assumptions! To put it another way, I will accept the idea of
subjective Bayesianism when this same subjectivity is acknowledged for
other methods of inference. Until that point, I prefer to speak not of
``subjectivity'' but of ``assumptions'' and ``scientific judgment.'' I
agree with Kass that scientists and statisticians can and should feel
free to make assumptions without falling into a ``solipsistic
quagmire.''

Finally, I am surprised to see Kass write that scientists believe that
the theoretical and real worlds are aligned. It is from acknowledging
the discrepancies between these worlds that we can (a) feel free to make
assumptions without being paralyzed by fear of making mistakes, and (b)
feel free to check the fit of our models (those hypothesis tests again!
Although I prefer graphical model checks, supplanted by $p$-values as
necessary). All models are false, etc.

I assume that Kass is using the word ``aligned'' in a loose sense, to
imply that scientists believe that their models are \textit{appropriate}
to reality even if not fully correct. But I would not even want to go
that far. Often\ in my own applied work I have used models that have
clear flaws, models that are at best ``phenomenological'' in the sense
of fitting the data rather than corresponding to underlying processes of
interest---and often such models do not fit the data so well
either.\footnote{In the annals of hack literature, it is sometimes said
that if you aim to write best-selling crap, all you will end up with is
crap. To truly produce best-selling crap, you have to have a~conviction,
perhaps misplaced, that your writing has integrity. Whether or not this
is a good generalization about writing, I have seen an analogous
phenomenon in statistics: If you try to do nothing but model the data,
you can be in for a wild and unpleasant ride: real data always seem to
have one more twist beyond our ability to model (von Neumann's
elephant's trunk notwithstanding). But if you model the underlying
process, sometimes your model can fit surprisingly well as well as
inviting openings for future research progress.} But these models can
still be useful: they are still a part of statistics and even a part of
science (to the extent that science includes data collection and
description as well as deep theories).

\section*{Different Schools of Statistics}

Like Kass, I believe that
philosophical debates can be a good thing, if they motivate us to think
carefully about our unexamined assumptions. Perhaps even the existence
of subfields that rarely communicate with each other has been a source
of progress in allowing different strands of research to be developed in
a pluralistic environment, in a way that might not have been so easily
done if statistical communication had been dominated by any single
intolerant group. Ideas of sampling, inference, and model checking are
important in many different statistical traditions and we are lucky to
have so many different ideas on which to draw for inspiration in our
applied and methodological research.

\section*{Acknowledgments}

This work was supported in part by Institute of Education
Sciences, Department of Energy, National Science Foundation and
National Security Agency.

\end{document}